\documentclass[aip,apl,reprint]{revtex4-1}
\usepackage{graphicx}
\usepackage{dcolumn}
\usepackage{bm}

\draft 

\begin{document}


\title{Narrowband fluorescent nanodiamonds produced from chemical vapor deposition films} 


\author{E. Neu}%

\author{C. Arend}%

\author{E. Gross}
\author{F. Guldner}

\author{C. Hepp}%

\author{D.Steinmetz}

\author{E. Zscherpel}%
\affiliation{Universit\"at des Saarlandes, Fachrichtung 7.2, 66123 Saarbr\"ucken, Germany}
\author{S. Ghodbane}%

\author{H. Sternschulte}%
\altaffiliation{present address: nanoTUM, Technische Universit\"at M\"unchen, 85748 Garching, Germany }
\author{D. Steinm\"uller-Nethl}%
\affiliation{rho-BeSt Coating Hartstoffbeschichtungs GmbH, 6020 Innsbruck, Austria}
\author{Y. Liang}%

\author{A. Krueger}%
 \email{anke.krueger@uni-wuerzburg.de}
\affiliation{Universit\"at W\"urzburg, Institut f\"ur Organische Chemie, 97074 W\"urzburg, Germany }
\author{C. Becher}%
 \email{christoph.becher@physik.uni-saarland.de}
\affiliation{Universit\"at des Saarlandes, Fachrichtung 7.2, 66123 Saarbr\"ucken, Germany}



\date{\today}

\begin{abstract}
We report on the production of nanodiamonds (NDs) with \mbox{70-80 nm} size via bead assisted sonic disintegration (BASD) of a polycrystalline chemical vapor deposition (CVD) film. The high crystalline quality NDs display intense narrowband (\mbox{7 nm}) room temperature luminescence at \mbox{738 nm} from in situ incorporated silicon vacancy (SiV) centers. We demonstrate bright, narrowband single photon emission with $>100000$ cps. Due to the narrow fluorescence bandwidth as well as the near-infrared emission these NDs are also suitable as fluorescence labels with significantly enhanced performance for in-vivo imaging.
\end{abstract}

\pacs{}

\maketitle 
\begin{figure}[t]
\includegraphics{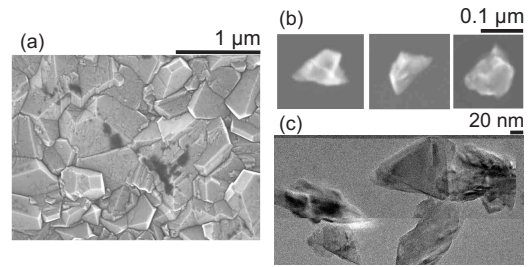}
\caption{\label{fig:remnd}SEM Images of: (a) the polycrystalline starting material, (b) individual NDs after spin coating onto a Si substrate. (c) High resolution transmission electron microscopy (HR-TEM) images of dropcast colloid on a copper grid with lacey carbon film. }
\end{figure}
In recent years fluorescent nanodiamonds (NDs) have attracted increasing attention in a wide field of prospective applications e.g. fluorescence labeling in biological imaging \cite{Chang2008,Mohan2010,Neugart2007} and as solid-state single photon sources (SPSs) \cite{Beveratos2002}. In these applications they stand out owing to photostable fluorescence \cite{Chang2008}, bio compatibility \cite{Mohan2010} and feasible surface functionalization \cite{Neugart2007}. The majority of the experiments performed in these fields rely on the fluorescence of the nitrogen vacancy (NV) center in diamond. NV centers are readily available as single centers in NDs\cite{Schietinger2009} or created by irradiating diamond powders \cite{Chang2008,Beveratos2002}. Recent work also demonstrated the efficient production of NV containing NDs by ball milling of microcrystalline powder \cite{Boudou2009}.  Despite these advantages, NV centers exhibit two major drawbacks for the applications mentioned above: their room temperature fluorescence spectrum spans about \mbox{100 nm} \cite{Schietinger2009} and, due to their zero-phonon-line (ZPL) at \mbox{637 nm}, excitation with visible laser light (typ. at \mbox{532 nm}) is necessary. On the other hand, the silicon vacancy (SiV) center delivers superior fluorescence properties. Firstly, it exhibits a ZPL at \mbox{$\approx$738 nm} \cite{Feng1993,Wang2006,Sternschulte1994}, lying within the near-infrared window of biological tissue\cite{Weissleder2003}. Furthermore, due to low electron-phonon-coupling about \mbox{80 \%} of the fluorescence is emitted into the ZPL even at room temperature\cite{Zaitsev2001,Neu2011} with an ensemble width of only \mbox{5-6.6 nm} \cite{Wang2006,Feng1993}. These properties lead to exceptional advantages in the potential applications of SiV-containing NDs: Firstly, the fluorescence can be excited using red laser light (here: 671 nm) thereby minimizing tissue autofluorescence for in-vivo imaging applications \cite{Chang2008,Weissleder2003}. Simultaneously, absorption of the excitation laser is $\approx$20 times lower compared to 532 nm\cite{Weissleder2003}, thus enabling imaging of deep tissue which is crucial for life science applications. Secondly, the significantly reduced fluorescence bandwidth allows for narrow spectral filtering, even further reducing autofluorescence in imaging applications. The narrow-band fluorescence also enables efficient discrimination of single photon emission and background signals for applications in e.g. quantum cryptography. As we show in this work, the SiV centers emit in a spectral region where background emission from the diamond material is low. Combined with the enhanced photon extraction from NDs\cite{Beveratos2002} SiV centers in NDs thus are promising candidates for practical solid state SPSs.

Commonly, SiV centers are observed in all types of chemical vapor deposition (CVD) diamonds including homoepitaxial single-crystalline \cite{Sternschulte1994} and polycrystalline materials \cite{Feng1993} due to Si incorporation by etching of Si substrates and CVD reactor walls. These SiV centers produced in situ during CVD growth seem to possess superior fluorescence properties as compared to SiV centers produced by ion implantation: the latter ones were examined as SPSs \cite{Wang2006} but showed unfavorably low emission rates (1000 counts per second (cps)). On the other hand, single SiV centers produced in situ in CVD grown NDs demonstrated very bright emission with \mbox{$10^6$ cps} recently \cite{Neu2011}. These substrate bound CVD NDs, however, are not applicable as fluorescence labels requiring NDs in solution or for enhanced SPSs which require spatial nanomanipulation to achieve coupling to photonic or nanoplasmonic structures\cite{Schietinger2009}.
We here report the production of fluorescent NDs containing in situ produced SiV centers from polycrystalline CVD films via the bead assisted sonic disintegration (BASD) method. These NDs combine the advantageous SiV fluorescence properties and its feasible production during the CVD process with the extended applicability of NDs dispersed in solution.

\begin{figure}
\includegraphics{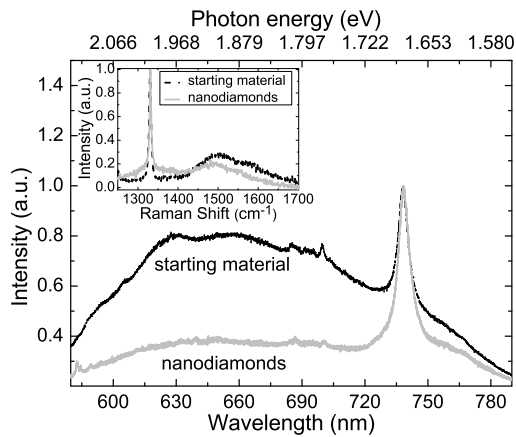}
\caption{\label{fig:plrmanens}Photoluminescence spectra recorded under 532 nm laser excitation from the polycrystalline starting material and an ensemble of the produced NDs. The inset shows the Raman spectrum of the starting material and the NDs recorded with 488 nm laser light.}
\end{figure}

As starting material for ND production we employ a polycrystalline diamond film grown by rho-BeSt Coating GmbH (Innsbruck) using hot filament CVD on ND seeded Si\cite{supplmat2011}. Scanning electron microscope (SEM) images indicate a film thickness of \mbox{2 $\mu$m} and high crystalline quality (grain size \mbox{0.5-1 $\mu$m}, Fig. \ref{fig:remnd}(a)). Raman spectroscopy reveals a distinct diamond Raman line (\mbox{1330.3 cm$^{-1}$}, inset Fig. \ref{fig:plrmanens}); its width of \mbox{5.9 cm$^{-1}$} indicates high crystalline quality with moderate stress distribution \cite{Zaitsev2001}. A weak G-band at around \mbox{1510 cm$^{-1}$} is attributed to residual sp$^2$ carbon. The film thus provides a high quality starting material for ND production. Photoluminescence measurements (Fig. \ref{fig:plrmanens}) display a pronounced SiV ZPL at \mbox{738.2 nm} (width \mbox{5.6 nm} in accordance with the literature \cite{Wang2006,Feng1993}). Additionally, we observe a broad band fluorescence that strongly diminishes beyond 700 nm\cite{supplmat2011}. The characterized diamond film is used for ND production after removing the substrate by chemical etching.

The BASD process, that some of us disclosed recently \cite{Ozawa2007,Liang2009}, enables the deagglomeration of strongly agglomerated nanoparticles by charging ceramic microbeads to a sonicated suspension. The microjets or shock waves induced by ultrasonic cavitations propel the beads (here: ZrO$_2$, \mbox{50 $\mu$m} size) leading to the disintegration of large agglomerates and the eventual formation of colloidal solutions of primary nanoparticles. So far, the method has only been applied for the deagglomeration of nanoparticles bound by inter-particle forces. Here, we apply  the BASD technique as a top down approach for the production of nanoparticles from a CVD diamond film by separating the grains and crushing the $\mu$m-sized crystals.  The BASD process is followed by several purification steps\cite{supplmat2011}, delivering a colloidal suspension of NDs in deionized water. The average particle size, determined by dynamic light scattering directly in the resulting solution, shows a distribution maximum at \mbox{70-80 nm}\cite{supplmat2011}. As the cumulative size distribution shows that more than \mbox{80 \%} of all particles have a size \mbox{$<$100 nm}\cite{supplmat2011}, these NDs are highly suitable for in-vivo imaging applications\cite{Mohan2010}. HRTEM and SEM observation verified the diamond nature of the nanoparticles as well as the nanometric size (Fig. \ref{fig:remnd}(b)$\backslash$(c)). The faceted shape of the individual NDs indicates the crushing of the film's crystallites along lattice planes. No contamination from the BASD process was found in the final colloidal solution\cite{supplmat2011}.
\begin{figure}
\includegraphics{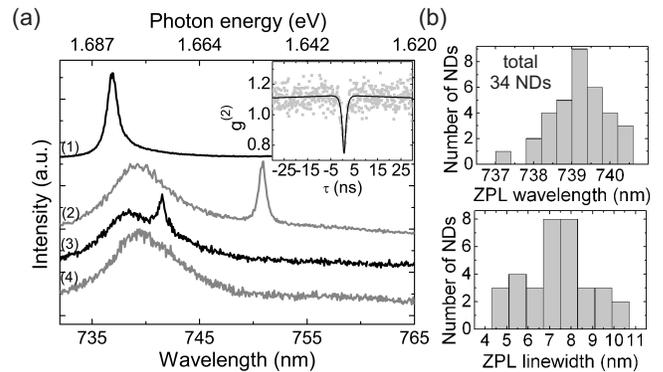}
\caption{\label{fig:ndhist}(a) Photoluminescence spectra of individual NDs. Inset: \mbox{g$^{(2)}$ function} of ND (1), fit taking into account background and instrument response revealing g$^{(2)}$(0)$\approx$0.05, (b) Histogram of ZPL positions and widths (mean value 6.8 nm). }
\end{figure}
\begin{figure}
\includegraphics{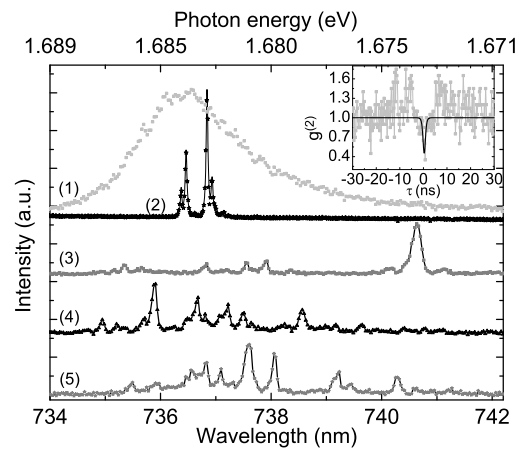}
\caption{\label{fig:ndcryo}Photoluminescence spectra at cryogenic temperatures. (1) SiV luminescence from 1.65 $\mu$m thick polycrystalline film at \mbox{4.7 K}, (2) SiV luminescence from high quality 100 nm thick homoepitaxial film at \mbox{23.5 K}, (3)-(5) spectra of individual NDs at \mbox{30 K}, Inset: \mbox{g$^{(2)}$ function} of ND (3) }
\end{figure}

To determine the fluorescence properties the ND solution is diluted and spin coated onto Si or Ir substrates. In a first step, we investigate a large ensemble of NDs on a sample with high density. Raman measurements certify conservation of the crystalline quality throughout the production process (Raman line \mbox{1331.2 cm $^{-1}$}, width \mbox{5.1 cm$^{-1}$}, inset Fig. \ref{fig:plrmanens}). Thus, the NDs provide high crystalline quality hosts for SiV centers, indispensable for achieving bright SiV luminescence\cite{Neu2011}.  Fig. \ref{fig:plrmanens} displays a typical fluorescence spectrum of a ND ensemble. We achieve significant reduction of broadband fluorescence, while the SiV luminescence remains basically unchanged (ZPL \mbox{738.6 nm}, width \mbox{6.8 nm}). The reduced background is very promising with respect to applications of the NDs as solid state SPSs. \\
To perform single ND spectroscopy on a low density sample we use a homebuilt confocal microscope setup \cite{Neu2011} (NA 0.8, excitation \mbox{660$\backslash$671 nm}). The spectra of four individual NDs are shown in Fig. \ref{fig:ndhist}(a), displaying intense SiV ZPLs. A histogram of the observed line positions is given in Fig. \ref{fig:ndhist}(b), displaying peak wavelengths between \mbox{737.4 nm} and \mbox{740.3 nm}. The 'ensemble ZPL' widths of 3.8 to \mbox{9.9 nm} confirm narrowband luminescence for all NDs investigated. The spread of the line positions and widths is caused by varying stress in the NDs\cite{Zaitsev2001,Neu2011}. We perform polarization dependent excitation studies, yielding up to \mbox{50 \%} visibility, evidencing (partial) alignment of the color centers inside the NDs. We point out that the observed fluorescence was photostable under excitation with up to \mbox{$\approx$1000 kW/cm$^2$}. The measured ensemble fluorescence rates did not saturate, partially exceeding the maximum count rate of our photon counters (\mbox{10$^7$ cps}). Therefore, the NDs containing larger ensembles of SiV centers are applicable as bright, narrowband, photostable fluorescence labels.

In addition to these ensemble ZPLs we observe lines as narrow as \mbox{0.8 nm} (see Fig. \ref{fig:ndhist}(a),(1)-(3)). Comparable line widths have been recently reported for single SiV centers \cite{Neu2011}, thus we tentatively attribute these lines to shifted, bright single SiV centers. We find that some of the NDs (Fig. \ref{fig:ndhist}(a),(1)) show dominant bright emission ($>100000$ cps) from a single SiV center, verified by nonclassical intensity autocorrelation \mbox{(g$^{(2)}$-)measurements}\cite{Beveratos2002,Neu2011} (Fig. \ref{fig:ndhist}(a) inset). We thus demonstrate  a SiV based SPS from a colloidal ND solution featuring a ZPL linewidth of $\approx$\mbox{1.5 nm} at room temperature.

To gain further insight into the ND fluorescence we perform low temperature (4.7 - 30 K) spectroscopy. Fig. \ref{fig:ndcryo} displays luminescence spectra from individual NDs, a high quality homoepitaxial and a polycrystalline CVD film. The observed lines blue-shift as compared to the room temperature ZPL in agreement with the literature\cite{Feng1993}. The spectrum of the homoepitaxial film displays a four line fine structure indicating a split excited and ground state \cite{Sternschulte1994}, while this structure is washed out in the polycrystalline film due to inhomogeneous broadening. The spectra of the NDs also show a fine structure (individual line widths \mbox{0.1 nm}),  but are more complex due to sub-ensembles of SiV centers under different stress, corresponding to the broadening in the polycrystalline film. Again some of the NDs show dominant emission from a single SiV center (Fig. \ref{fig:ndcryo} inset), evidencing low temperature single photon emission.

\begin{figure}[h!!!!!!!!!!!!!!!!!!!!!!!!!!!!!!!!!!!!!!!!!!!!!!!!!!!!!!!!!!!!!!!!!!!!!!!!!!!!]
\includegraphics[width=0.35\textwidth]{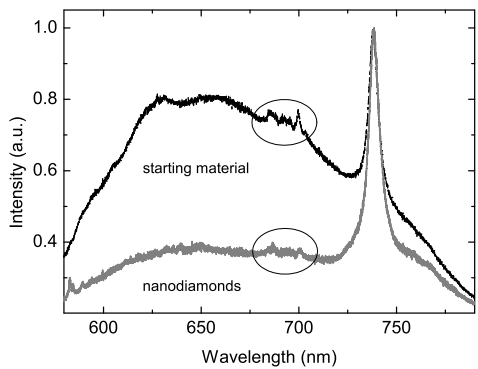}
\caption{\label{fig:plrmanens_suppl}Photoluminescence spectra recorded under 532 nm laser excitation from the polycrystalline starting material and an ensemble of the produced NDs. The marked areas show the tantalum related lines.}
\end{figure}

In summary we have demonstrated that BASD of diamond films produces fluorescent NDs that preserve or even enhance favorable properties of the starting material i.e. high brightness, narrowband fluorescence, low background emission and high photostability. We demonstrate bright ($>100000$ cps) single photon emission from SiV centers at room temperature. The fluorescent NDs also offer great potential as optimized fluorescence labels, where additional Si doping could further enhance the performance. For single photon applications, smaller NDs produced from purer diamond material are desirable. As the method presented here is easily scalable it pioneers the production of large amounts of NDs with engineered defect properties from a variety of diamond films.
\begin{acknowledgments}
We acknowledge funding by the DFG, the European Commission (EQUIND, DINAMO, DRIVE) and the BMBF (EPHQUAM 01BL0903).
SEM measurements were performed by J. Schmauch (UdS).
\end{acknowledgments}
\newpage
\textbf{\large Supplementary material for: \\Narrowband fluorescent nanodiamonds produced from chemical vapor deposition films\\
}
\textit{The supplementary material covers details of the BASD process, including the employed purification steps, as well as the data of the nanodiamond (ND) size measurements by dynamic light scattering (DLS). We give additional information on the luminescence spectra discussed in the paper.}\\
\noindent \textbf{Additional discussion on fluorescence spectra and starting material:}\\
The polycrystalline diamond film used as a starting material has been grown by rho-BeSt using a hot filament CVD Process. As substrate a 4' ND seeded \{100\}-oriented Si wafer has been used. The growth was performed using \mbox{0.26 \%} CH$_4$ in H$_2$. Taking into account the SEM images, the film provided approx. 0.053 g of diamond material.\\
In Fig. \ref{fig:plrmanens_suppl} (cf. Fig 2, original manuscript) additional lines between 680 and \mbox{705 nm} are observed in the fluorescence spectrum.
As these lines are not relevant for the SiV related topics they are not discussed in the paper. This triplet has been tentatively attributed to the incorporation of tantalum from the hot filament process \cite{Harris1996}. As these fluorescence lines are rather weak and nearly coincide with the maximum of the broad fluorescence background observed, the application of these centers as fluorescence markers or single photon sources is unfavorable.\\
\textbf{Additional information on the BASD process:}\\
\begin{figure}[t!!!!!!!!!!!!!!!!!!!!!!!!!!!!]
\includegraphics[width=0.5\textwidth]{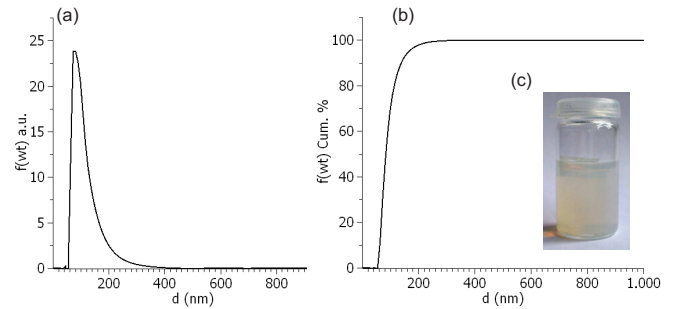}
\caption{\label{fig:suppl_DLS_sol}(a) size distribution of the ND colloid, (b) cumulative size distribution of the colloidal solution. (c) ND colloid in deionized water obtained from a BASD-treated CVD diamond film}
\end{figure}
As described in the manuscript we apply  a BASD process to crush a polycrystalline diamond film. So far, the method has only been applied for the deagglomeration of nanoparticles bound by inter-particle forces (electrostatic, $\pi$-$\pi$ interactions, hydrogen bonding, bonds between surface groups). Here, the BASD technique has been applied for the production of nanoparticles from a CVD diamond film, as these films are not easily processed in a conventional grinding tool such as a mortar or ball mill due to technological restrictions (available amounts of starting material, applicable forces, necessity to crush the crystallites along crystal planes etc.).
In preliminary experiments to this study we observed that beadless ultrasonic treatment was not effective for the crushing of diamond films. Therefore, the BASD process and the following purification have been carried out: Sheets of the diamond film and ZrO$_2$ microbeads (50 $\mu$m) were treated in \mbox{5 ml} of dimethyl sulfoxide (DMSO) for \mbox{2 h} using a powerful sonicator (\mbox{400 W}) equipped with a horn-type sonotrode. \mbox{200 ml} of toluene were added to the reaction mixture. The sediments were separated by centrifugation and washed with \mbox{100 ml} of acetone. In order to remove impurities like sonotrode abrasion, amorphous carbon and nanozirconia fragments, the residue was first stirred overnight in a 1:1:1 mixture of conc. H$_2$SO$_4$, HNO$_3$ and HClO$_4$ at \mbox{85 $^\circ$C}. Secondly, the zirconia beads have been removed by centrifugations. The resulting supernatant was then stirred in \mbox{20 N KOH} overnight and centrifuged. The sediment was washed in repeated centrifugation-redispersion cycles with deionized water until \mbox{pH 7} and stored as a colloidal solution (Fig. \ref{fig:suppl_DLS_sol}(c)) in deionized water. The average particle size, i.e. hydrodynamic diameter determined by dynamic light scattering (DLS) directly in the colloidal solution of the resulting ND particles, is well below \mbox{100 nm} with the distribution maximum at \mbox{70-80 nm} (Fig. \ref{fig:suppl_DLS_sol}(a)) . The cumulative size distribution shows that more than \mbox{80 \%} of all particles have a size \mbox{$<$ 100 nm} (Fig. \ref{fig:suppl_DLS_sol}(b)). EDX measurements after the purification proved the absence of zirconia contamination.

\end{document}